\RequirePackage{fix-cm}
\documentclass[twocolumn,epjc3]{svjour3}  
\smartqed  
\RequirePackage{graphicx}
\usepackage{bm}
\usepackage{amssymb}
\usepackage{amsbsy}
\usepackage{physics}
\usepackage{xcolor}

\journalname{Eur. Phys. J. C}
\begin{document}

\title{Prospecting
Black Hole Thermodynamics 
with Fractional Quantum Mechanics}


\author{S. Jalalzadeh\thanksref{e1,addr1}
        \and
        F. Rodrigues da Silva\thanksref{e2,addr1} \and
        P.V. Moniz \thanksref{e3,addr2,addr3}
}

\thankstext{e1}{e-mail: shahram.jalalzadeh@ufpe.br}
\thankstext{e2}{e-mail: filipe.rodrigues@ufpe.br}
\thankstext{e3}{e-mail: pmoniz@ubi.pt}

\institute{Departamento de F\'{i}sica, Universidade Federal de Pernambuco,
Recife, PE, 52171-900, Brazil \label{addr1}
           \and
           Departmento de F\'{i}sica , Universidade da Beira Interior, 6200 Covilh\~a, Portugal \label{addr2}
           \and
           Centro de Matem\'atica e Aplica\c c\~oes (CMA-UBI), Covilh\~a, Portugal\label{addr3}
}

\date{Received: date / Accepted: date}
\maketitle
\begin{abstract}
This paper investigates whether the fra\-me\-work of fractional quantum mechanics can broaden our perspective of black hole thermodynamics. Concretely, we employ a {\it space-fractional} deri\-vative \cite{Rie} as our main tool. Moreover, we restrict our analysis to the case of a Schwarzschild configuration. From a subsequently modified Wheeler-DeWitt equation, we retrieve the corresponding expressions for specific observables. Namely, the black hole mass spectrum, $M$, its temperature $T$, and entropy, $S$. We find that these bear consequential alterations 
conveyed through a fractional parameter, $\alpha$. In particular, 
the standard results are recovered in the specific limit $\alpha=2$. 
Furthermore,  we elaborate how generalizations of the  entropy-area relation suggested by Tsallis  and Cirto \cite{Tsallis} and Barrow \cite{Barrow} acquire a complementary interpretation in terms of a fractional point of view. A thorough discussion of our results is presented.
\end{abstract}

\section{Introduction}\label{S1}

\indent

Black hole (BH) physics constitutes an enthusiastic research domain for this century.
On the one hand, gravitational waves resonating from colliding BHs 
 have been measured \cite{Abbott} and are now a regular astronomical probing tool. Outstandingly, astrophysicists have also captured the first-ever image of a BH at the center of galaxy M87 \cite{Aki}. Furthermore, there is a collection of additional observational data \cite{Event} hinting at the presence of an  event horizon for BHs. 
In general, BH  candidates may be   stellar-mass objects in a X-ray binary system 
or 
supermassive  at the center of typical galaxies \cite{Ritter,Rees,Stellar}. All these achievements 
have been recently conveyed and recognized within the 2020 Nobel prize in Physics \cite{Nobel}.

On the other hand, innovative work  on BH physics in the past forty
years or so have  also brought us to the forefront of  unanticipated  research directions. In the early 70's
of the last century, 
Bekenstein proposed that a BH entropy is proportional to its horizon area \cite{Bec}. Then, the evaporation of BHs (through the seminal use of quantum fields in curved spaces) was   formulated 
by Hawking \cite{Hawking1}. {Testing this property observationally has been recently appraised, bearing promising results \cite{GW-BH-S}.} Thereafter, physicists realized an intimate connection between geometrical horizons, thermodynamic temperature and quantum mechanics \cite{Fulling}. Moreover, it was advanced \cite{Bekenstein} that the BH horizon area may be  quantized, and the corresponding eigenvalues would be given by 
\begin{eqnarray}
\label{AA1}
A_n=\gamma L_\text{P}^2n,~~~~n=1,2,3,...~,
\end{eqnarray}
where $\gamma$ is a dimensionless constant of  order one and $L_\text{P}=\sqrt{G}$ is the Planck length. The literature has since then been enriched with contributions strengthening  in favor of the area spectrum (\ref{AA1}), including information-theoretic considerations \cite{Area1}, ranging  from string theory arguments \cite{Area2}  to  the  periodicity of time \cite{Area3}, plus e.g., a Hamiltonian quantization of a dust collapse \cite{Area4}.


{Within the viewpoint  outlined in the previous paragraph,  we suggest a  complementary direction of exploration.
Let us be more specific. Recently, a set of pertinent arguments and results 
have been put forward   to apply  fractional calculus \cite{par1,par2,par3,par4} in quantum physics. Such framework is known as Fractional Quantum Mechanics (FQM), see e.g. \cite{Arxiv}.} 
{Fractional calculus 
has been  embraced mainly  within the last century.
In essence, 
it follows from extending  the meaning of derivatives to  the case where the order is any number, i.e., irrational, fractional or complex. {Its peculiar interest notwithstanding, 
the obstacles have been fruitfully addressed. 
In particular, it is currently  known that}  non-integer order systems can describe the dynamical behavior of 
specific classes of materials and processes over 
different 
time and frequency scales.
Fractional calculus has assisted in  scattering theory, diffusion, 
probability, potential theory and elasticity. 
Therefore, it was only sensible to embrace and 
explore it within quantum physics.}

{FQM was built upon Feynman and Hibbs \cite{Bar1} assertion about quantum mechanics and path integrals. 
Moreover, Nelson \cite{Bar2}  clarified  
about classical Brownian motion and quantum mechanical features, whereas  Abbott and Wise \cite{Bar3} determined that in $1$-dimensional settings, the fractal (or Hausdorff) dimension of those paths is 2. The essential motivation for FQM emerges in that if we restrict the path integral (Feynman) description of quantum mechanics to 
 Brownian paths only, it will be challenging to explain a few  other pertinent quantum phenomena \cite{Laskin1}. Such difficulties have  led  to consider  a generalization of the Feynman path integral, specifically  by replacing the Gaussian probability distribution by L\'evy's \cite{Levy};  The Hausdorff dimension of the L\'evy path is then equal to the fractional parameter $\alpha$. 
 }


{Extended  versions of the Schr\"odinger equation (SE) are then retrieved, namely from such broader path integral. More concretely, by including non-Brownian trajectories in the path integral formalism of quantum physics, either the space-fractional \cite{space}, time-fractional \cite{time}, and space-time-fractional \cite{space-time} versions of the standard SE can be elaborated. Let us be precise and clarify that it is in space-fractional quantum mechanics, where  the Feynman path integral method is modified  such that the Gaussian probability distribution is replaced by L\'evy's \cite{Levy},   that we obtain the following and  broadly used  modified SE.}
{More particularly,
starting from the  Hamiltonian 
\begin{eqnarray}
\label{2ref1}
H=\frac{\mathbf p^2}{2m}+V(\mathbf r,t),
\end{eqnarray}
a generalization \cite{Arxiv,Laskin1,space}
is produced by means of 
\begin{eqnarray}
\label{2ref2}
H_\alpha
(\mathbf{p}, \mathbf{r}) : = D_\alpha
|\mathbf{p}|^\alpha
+ V (\mathbf{r}), \hspace{10mm} 1 < \alpha \leq  2,
\end{eqnarray}
in which $D_\alpha$ is a coefficient carrying
dimension $[D_\alpha]={\rm erg}^{1-\alpha}{\rm cm}^\alpha{\rm sec}^{-\alpha}$. }
{The parameter $\alpha$ is known as L\'evy's fractional parameter and is associated to the concept of  L\'evy path and its induced fractal dimension. In the Feynman path integral, the measure is generated by the process of the Brownian motion, and the path's corresponding (fractal) dimension  is $d^{\text{(Feynman)}}_\text{fractal}=2$, as 
we have remarked. The L\'evy path integral leading to a SE would lead to a fractal dimension  
as $d_\text{fractal}^{\text{(L\'evy)}}=\alpha$. For more 
details, please browse through, e.g.,  \cite{Rie,par1,par2,par3,Arxiv,Laskin1,space} and other references therein.}

{Therefore, choosing a space representation
with  ${\hat{\mathbf{p}}}\rightarrow -i\hbar \nabla$
and ${\hat{\mathbf{r}}}\rightarrow {\mathbf{r}}$,} we obtain the space-fractional SE
\begin{equation}
    i\hbar \frac{\partial \psi(\mathbf{r},t)}{\partial  t}=
D_\alpha
(-\hbar^2 \Delta)^{\alpha/2} \psi(\mathbf{r},t)
+ V(\mathbf{r},t) \psi(\mathbf{r},t).
\label{7-0}
\end{equation}
In the above fractional SE, the
 fractional  Riesz derivative \cite{Rie,Arxiv,Kilbas}, $(-\hbar^2 \Delta)^{\alpha/2}$, is defined in terms of the Fourier transformation $\mathcal F$
 \begin{equation}
 \begin{array}{cc}
(-\hbar^2 \Delta)^{\alpha/2} \psi(\mathbf{r},t)
=\mathcal F^{-1}|\mathbf{p}|^\alpha\mathcal F\psi(\mathbf{r},t)\\
\\
=\displaystyle\frac{1}{(2\pi\hbar)^3}
\int d^3 p e^{i\frac{\mathbf{p}\cdot \mathbf{r}}{\hbar}}
|\mathbf{p}|^\alpha
\int e^{-i\frac{\mathbf{p}\cdot \mathbf{r}'}{\hbar}}\psi(\mathbf{r}',t)d^3r'.
\label{2ref8}
\end{array}
\end{equation}
The infinite-well  example was one of the first solutions of space-fractional SE, which Laskin solved \cite{Laskin1}. Despite its simplicity, this problem is critical since it is the prototype of a quantum detector with internal degrees of freedom.

{In what concerns  the time-fractional  SE, the time evolution is given by the Caputo fractional derivative \cite{Caputo}. 
In this case, the associated Hamiltonian is non-Hermitian and not local in time.
The space-time fractional SE has been introduced by Wang and Xu \cite{space-time}, where they employed a combination of space and time-fractional models to establish their equation.}
{The space-time fractional SE is a generalized version of the equation \eqref{7-0}:
\begin{equation}
   \hbar_\beta i^\beta \partial_t^\beta \psi(\mathbf{r},t)
= \left[D_{\alpha,\beta}(-\hbar_\beta^2 \Delta)^{\alpha/2}+V(\mathbf{r},t)\right]\psi(\mathbf{r},t),   
\label{ST-1}
\end{equation}
where $1<\alpha\leq2, ~0<\beta \leq1$,
 $ \hbar_\beta$ and $D_{\alpha,\beta}$ are
 two scale coefficients with physical dimensions $[\hbar_\beta]={\rm erg}.{\rm sec}^\beta$ and
  $[D_{\alpha,\beta}]={\rm erg}^{1-\alpha}.{\rm cm}^\alpha.{\rm sec}^{-\alpha\beta}$. In addition, $\partial_t^\beta$
denotes the left Caputo fractional derivative \cite{C67} of order $\beta$:
\begin{equation}
 \partial_t^\beta f(t)=\frac{1}{\Gamma(1-\beta)}\int_{0}^{t}d\tau\frac{\dot f(\tau)}{(t-\tau)^\beta},
\label{ST-3}
\end{equation}
where $\dot f(\tau)= \frac{df(\tau)}{d\tau}$.
}

{Proceeding towards a broader context, we recall that  Wheeler \cite{Bar4} seminally  suggested a foam structure for spacetime on the Planck scale.  Hence, allowing that a fractal quality could therein be natural.
Thus, contemplating a fractional WDW equation could unveil interesting aspects regarding the gravitational domain \cite{FQ2,FQ3}. In fact, 
in the last few years, FQM has indeed been 
proposed and employed as a tool to explore features within quantum cosmology and quantum gravity \cite{FQ1}.  It has pointed to exciting opportunities and sidelong connections between unexpected mathematics and physics domains. See  \cite{FQ2,FQ3}, for a recent  survey about engaging  FQM  within a  canonical route   towards quantum gravity and cosmology. Additionally, a few more other references can be found in \cite{all-new}.}

Thus, in our paper we propose to apply  space-FQM to investigate 
thermodynamic properties of the Schwarz\-schild BH. The Bekenstein-Hawking entropy expression for a Schwarzchild BH with mass $M$ is given by 
\begin{eqnarray}\label{BH}
S_\text{B-H}=4\pi GM^2=\frac{A}{4G},
\end{eqnarray}
where $A=16\pi G^2M^2$ is the BH horizon area. A suitably extended fractional version of the Wheeler-DeWitt (WDW) equation for the Schwarzschild BH will enable us to  {retrieve 
altered expressions for the entropy and other specific BH observables. }

Moreover, Tsallis and Cirto recently proposed a new formulation for the Schwarzschild BH's horizon entropy and the entropy-area relation \cite{Tsallis}.  They asserted that since the Bekenstein-Hawking entropy is proportional to the  horizon area 
 (instead of its volume), it would hint that Boltzmann-Gibbs entropy would be unsuitable to describe BHs. These authors have proposed a modified form upon the  BH entropy given by Eq. (\ref{BH}),  such that 
 an extended  expression,  using   a generalized non-additive entropy, would  be
\begin{eqnarray}\label{Ts1}
S_\delta=\gamma A^\delta,
\end{eqnarray}
where $\gamma$ is an unspecified constant and $\delta$ denotes the non-additivity parameter. The above equation suggests that the  entropy  is a power-law function of its area. 

In addition, by introducing a fractal structure for the horizon surface of the BH, 
Barrow \cite{Barrow}  
{conceived a 3-dimensional spherical analog of a Koch Snowflake--a sphereflake-- using a hierarchy of touching spheres around the event horizon. This fractal} structure of the BH surface changes its actual area, which in turn leads us to a new entropy relation, namely,
\begin{eqnarray}
S=\left(\frac{A}{A_P}\right)^{1+\frac{\triangle}{2}},
\end{eqnarray}
where $A_P$ is the Planck area and $\triangle$ 
{would suggestively denote an
induced deformation of the horizon}, with $\triangle=0$ reproducing the conventional Bekenstein-Hawking entropy (simplest horizon structure) and with $\triangle=1$ corresponding to the most intricate structure. Note that the Barrow 
modified entropy resembles Tsallis and Cirto's non-additive entropy (\ref{Ts1}); {nevertheless}, the involved foundations and physical principles are entirely different.
As an additional  application of  BH prospecting  by means of FQM, we {will retrieve these} 
two extended  entropy expressions but  without using generalized non-additive features.

{In essence, we propose  and employ herewith a  space-fractional formulation of the WDW equation for the Schwarzschild BH.}
Hence, let us {then} mention that the remaining of this paper is organized as follows. In section 3, we 
{summarize the} 
canonical quantization of the Schwarz\-schild BH, 
having briefly reviewed  the corresponding  Hamiltonian setting in section 2.  In Section 4, we hypothesize a fractional calculus setting  for the 
{corresponding WDW equation as applied to the}
Schwarz\-schild BH. Specifically, we consider the {observables reviewed}  in section 3 and present 
{them 
(e.g.,  }the Schwarz\-schild BH  entropy) from within the framework of FQM. Section {5} contains {our conclusions and  a} thorough  discussion. Throughout  this paper we shall work in natural units, $\hbar=c = k_B = 1$.

\section{Canonical quantization of a Schwarzchild black hole}\label{S2}

\indent 

We recall in this section 
results concerning the ca\-noni\-cal quantization of the Schwarzschild BH that  will be of relevance for our analysis in subsequent sections. The spherical symmetric ADM line element is
\begin{multline}\label{1-1}
ds^2=-N(r,t)^2dt^2+
\\\Lambda(r,t)^2\Big(dr+N^r(r,t)dt\Big)^2+R(r,t)^2d\Omega^2,
\end{multline}
where $d\Omega^2$ is the line element for the unit two sphere $S^2$. We follow Kucha\v r's fall-off conditions \cite{Kuchar1}.
Firstly, it confirms that the coordinates $r$ and $t$ are extended to the Kruskal manifold, $-\infty<r,t<\infty$, and secondly, that the spacetime is asymptotically flat as well. Likewise, the fall-off conditions guarantee that the 4-momentum at the infinities, $r\rightarrow\pm\infty$, has no spatial component, which indicates that the 
BH is at rest concerning the left and right asymptotic Minkowski spacetimes.
{By fixing the asymptotic values of the lapse function, $N$, at infinities, $r\rightarrow\pm\infty$ to be $t$-dependent quantities (denoted by $N_\pm(t)$, respectively), the Hamiltonian form of the Einstein-Hilbert action functional, with appropriate boundary terms, reads then}
\begin{multline}\label{1-2}
\mathcal S=\displaystyle\int dt\int_{-\infty}^{\infty}\Big\{ \Pi_\Lambda \dot\Lambda +\Pi_R\dot R-NH-N^rH_r\Big\}dr\\ -\displaystyle\int\Big\{N_+M_++N_-M_-\Big\}dt,
\end{multline}
in which the conjugate momenta of $\Lambda$ and $R$ are 
\begin{equation}\label{1-3}
\begin{split}
\Pi_\Lambda&=-\frac{M^2_P}{N}R\Big(\dot R-R'N^r\Big),\\
\Pi_R&=-\frac{M_P^2}{N}\Big[\Lambda\left(\dot R-R'N^r\right)+R\left(\dot\Lambda-(\Lambda N^r)'\right)\Big],
\end{split}
\end{equation}
 $M_P=1/\sqrt{G}$ is the Planck mass, $\dot f=\partial_t f$, $f'=\partial_rf$ and {the quantities $M_{\pm}(t)$ are defined by the asymptotic fall-off of the configuration variables. Note that on a classical solution, $M_\pm$ are equal to the Schwarzschild mass of the BH}. 
{Furthermore, $H$ and $H_r$ are the super-Hamiltonian and the radial super-momentum constraints, respectively, given by
\begin{equation}\label{new1}
\begin{split}
H=&-\frac{1}{RM_P^2}\Pi_R\Pi_\Lambda+\frac{1}{2R^2M_P^2}\Pi_\Lambda^2+\frac{RR''}{\Lambda}\\
&-\frac{RR'\Lambda'}{\Lambda^2}+\frac{R'^2}{2\Lambda}-\frac{\Lambda}{2},\\
 H_r&=\frac{1}{M_P^2}(\Pi_RR'-\Lambda\Pi'_\Lambda). 
 \end{split}
\end{equation}
Following Kucha\v r  \cite{Kuchar1} introducing the following two $(M,\Pi_M)$ and $(\mathcal R,\Pi_{\mathcal R})$ pairs of canonical transformations
\begin{equation}\label{new2}
\begin{split}
    M&=\frac{\Pi_\Lambda^2}{2M_P^4R}-\frac{RR'^2}{2\Lambda^2}+\frac{R}{2},  \\
    \Pi_M&=\frac{\Lambda\Pi_\Lambda}{M_P^2}\left[\left(\frac{R'}{\Lambda}\right)^2-\frac{1}{M_P^4}\left(\frac{\Pi_\Lambda}{R}\right)^2 \right]^{-1},\\
    \mathcal R&=R,\\
    \Pi_{\mathcal R}&=\left(\frac{\Pi_\Lambda H}{M^2_PR}+\frac{R'H_r}{\Lambda^2} \right)\left[\left(\frac{R'}{\Lambda}\right)^2-\frac{1}{M_P^4}\left(\frac{\Pi_\Lambda}{R}\right)^2 \right]^{-1},
\end{split}
\end{equation}
the action (\ref{1-2}) becomes
\begin{multline}\label{new3}
\mathcal S=\displaystyle\int dt\int_{-\infty}^\infty\left\{ \dot M\Pi_M+\dot{\mathcal R}\Pi_{\mathcal R}-N^rH_r-NH\right\}dr\\
-\displaystyle\int \left\{M_+N_+-M_-N_-\right\}dt,
\end{multline}
where the new super-Hamiltonian, $H$, and the super-momentum, $H_r$, are
\begin{multline}
\label{new4}
    H=\\-\frac{\left(1-\frac{2M}{M_P^2R}\right)^{-1}M'{\mathcal R}'+M_P^{-4}\left(1-\frac{2M}{M_P^2R}\right)\Pi_M\Pi_{\mathcal R}}{\left[\left(1-\frac{2M}{M_P^2R}\right)^{-1}{\mathcal R}'^2-M_P^{-4}\left(1-\frac{2M}{M_P^2R}\right)\Pi_M^2 \right]^\frac{1}{2}},  
     \\
     H_r=\frac{1}{M_P^2}\left(\Pi_MM'+\Pi_{\mathcal R}{\mathcal R}'\right).
\end{multline}
The variation of ADM action (\ref{new4}) with respect to the lapse function $N$ and
$N^r$ imposes the Hamiltonian and momentum constraints
\begin{eqnarray}
\label{new5}
H\approx0,~~~~H_r\approx0,
\end{eqnarray}
or equivalently
\begin{eqnarray}
\label{new6}
M'\approx0,~~~~\Pi_{\mathcal R}\approx0.
\end{eqnarray}
The constraint $M'= 0$ means that $M$ is homogeneous $M=M(t)$. Now, by substituting $\Pi_{\mathcal R}\approx0$ and $M=M(t)$ back into the action (\ref{new4}) and in addition the new conjugate momenta of $M$ defined by
\begin{multline}\label{1-4}
P=\displaystyle\int_{-\infty}^\infty \Pi_Mdr=
\\-\displaystyle\int_{-\infty}^\infty\frac{\sqrt{\left(\frac{dR}{dr}\right)^2-\Lambda\left(1-\frac{2M}{M^2_PR}\right)}}{1-\frac{2M}{M^2_PR}}dr,~~~ -\infty<P<\infty, 
\end{multline}
we obtain
}
\begin{eqnarray}\label{1-5}
\mathcal S=\int \Big\{P\dot M-(N_++N_-)M\Big\}dt.
\end{eqnarray}
Note that the new conjugate variables $(M, P)$ obey the Poisson bracket $\{M, P\}=1$. Following Louko and M\"akel\"a \cite{Louko}, if we picked the right-hand side asymptotic Minkowski time as the observer time parameter, we should restrict $N_+=1$ and $N_-=0$.  Next, the reduced action (\ref{1-5}) reduces to the following simple form
\begin{eqnarray}\label{1-6}
\mathcal S=\int \left\{P\dot M-H(M)\right\}dt,
\end{eqnarray}
where $H(M)=M$ is the reduced Hamiltonian of the BH. 
One can easily show that the field equations' solution is $M=const.$ and $P=-t$, as expected. 
The BH's mass constancy follows Birkhoff's theorem, which declares that the mass is the only time-independent and coordinate invariant solution. Moreover, the conjugate momenta, $P$, describes the asymptotic time coordinate at the spacelike slice. Once again, by using the canonical transformations $(M,P) \rightarrow (x,p)$,  introduced by Louko 
{and M\"akel\"a} \cite{Louko}
\begin{equation}\label{1-7}
\begin{split}
|P|&=\displaystyle\int_x^{2MG}\frac{dy}{\sqrt{\frac{2MG}{y}-1}},\\
M&=\frac{1}{2G}\left(\frac{G^2p^2}{x}+x\right),
\end{split}
\end{equation}
 the reduced action (\ref{1-6}) takes the form
\begin{eqnarray}\label{1-8}
\mathcal S=\int\left\{p\dot x-H\right\}dt,
\end{eqnarray}
where $H=M$ is the Hamiltonian and is given (utilizing the second transformation of (\ref{1-7})), as 
\begin{eqnarray}\label{1-9}
H=\frac{M_P^2}{2}\Big(\frac{p^2}{M_P^4x}+x\Big).
\end{eqnarray}
{It is of interest to acknowledge the following at this point. } 
Transformations (\ref{1-7}) map the 
BH solution into a
wormhole solution,  in which $x$ represents the wormhole throat \cite{Louko}. The time evolution of this corresponding wormhole throat is given by Hamilton's equations
\begin{eqnarray}\label{1-10}
\dot x=\frac{p}{M_P^2x},~~~~
\dot p=\frac{p^2}{2M_P^2x^2}-\frac{1}{2}M_P^2
\end{eqnarray}
with a solution for the wormhole throat as
\begin{equation}\label{1-11}
\begin{split}
x&=\frac{M}{M_P^2}\Big(1+\cos(M_P\eta)\Big),\hspace{.5cm} x\geq0\\
t&=\frac{M}{M_P}\Big(\eta+\frac{1}{M_P}\sin(M_P\eta)\Big),\hspace{.5cm} -\frac{M\pi}{M_P^2}\leq t\leq \frac{M\pi}{M_P^2}.
\end{split}
\end{equation}
As we 
can see from the above solutions, it is necessary to define transformations (\ref{1-7}) 
so that the time parameter $t$ (or equivalently the conjugate momentum $P$) is restricted into the finite interval
\begin{eqnarray}\label{1-12}
-\frac{1}{8T_H}\leq t\leq \frac{1}{8T_H},\hspace{0.5cm}T_H=\frac{M^2_P}{8\pi M},
\end{eqnarray}
where $T_H$ is, {remarkably},  the Hawking temperature of the BH.

\section{Thermodynamic implications from the quantum Schwarzschild black hole}
\label{S3}

\indent 

{In the coordinate representation $\hat p \rightarrow -id/dx$, $\hat x \rightarrow x$,} the canonical quantization procedure {upon} the previous section gives us a suitable  time-independent WDW equation, for the simple one-dimensional minisuperspace  of the  Schwarzschild BH
\begin{eqnarray}\label{2-1}
H\left(-i\frac{d}{dx},x\right)\psi(x)=M\psi(x).
\end{eqnarray}
{As usual, there is an operator-ordering problem in addressing the above WDW equation, but as we are interested in  BH  states whose mass $M \gg M_P$, its particular resolution will not {surpass
our 
semiclassical considerations. }
{Hence, we} adopt the following wide enough factor-ordering \cite{Jalalzadeh}
\begin{eqnarray}
\frac{p^2}{x}=\frac{1}{3}\left(x^ipx^jpx^k+x^kpx^ipx^j+x^jpx^kpx^i\right),
\label{j-order}
\end{eqnarray}
where $i+j+k=-1$. 
{Note that because in the left hand side of (\ref{j-order})  we have $p^2/x$, so, in its right side   the power of $x$ is also $-1$}.} We thus write the WDW equation (\ref{2-1}) 
\begin{multline}\label{new21}
    -\frac{1}{2M_P^2x}\frac{d^2}{d^2}\psi(x)+\frac{1}{2M_P^2x^2}\frac{d}{dx}\psi(x)+
\\\left(\frac{1}{2}M_P^2x+\frac{q}{2M_P^2x^3}\right)\psi(x)=M\psi(x),
\end{multline}
where $q=(ij+ik+jk-2)/3$. If we redefine the wave function $\psi(x)\rightarrow\sqrt{x}\psi(x)$, and choose ordering in which $q=-3/4$, then Eq. (\ref{new21}) will reduce to
\begin{multline}\label{2-2}
-\frac{1}{2M_P}\frac{d^2}{dx^2}\psi(x)+\frac{1}{2}M_P\omega_P^2\left(x-\frac{M}{M_P^2}\right)^2\psi(x)=\\\frac{M^2}{2M_P}\psi(x),
\end{multline}
which is a SE for the harmonic oscillator of the Planck's mass $M_P$, the Planck's angular frequency $\omega_P=1/t_P$ defined in terms of Planck's time $t_P=1/M_P$. 
The domain of definition for $x$ is $x\geq0$, and consequently, the Hamiltonian operator of the harmonic oscillator  in (\ref{2-2}) is defined on a dense domain $C^\infty(0,+\infty)$. Hence, $H$ {is not an essentially self-adjoint operator}. 
It may constitute an Hermitian operator if
\begin{eqnarray}\label{2-3}
\langle\psi_1|H\psi_2\rangle=\langle H\psi_1|\psi_2\rangle,\hspace{0.5cm}\psi_1,\psi_2\in {\mathcal D}(H).
\end{eqnarray}
A necessary and sufficient condition for {the} validity of this condition is
\begin{eqnarray}\label{2-4}
\left.\frac{\psi(x)}{\psi'(x)}\right\vert_{x=0}=\gamma,\,\,\,\,\gamma\in\mathbb R,
\end{eqnarray}
where a prime symbol, $\prime$, denotes the derivative of $\psi( x )$ with respect to $x$. As pointed out by Tipler \cite{Tipler},
 the constant $\gamma$ (with dimension of length) would be a new fundamental constant of theory. To avoid such, 
 we set it to be zero. Hence, we assume
 \begin{eqnarray}\label{2-4a}
 \psi(x)\Big|_{x=0}=0,
 \end{eqnarray}
which constitutes
the 
DeWitt boundary condition \cite{DeWitt}. In addition,  
we are interested in {square}-integrable
 wave functions
in the interval $0\leq x<+\infty$,  which implies the second boundary condition $\psi(x\rightarrow+\infty)=0$.
The general solution of the WDW equation (\ref{2-2}) is 
\begin{multline}\label{2-5}
\psi(z)=e^{-\frac{z^2}{4}}\Big\{A\, _1F_1\left(-\frac{\nu}{2};\frac{1}{2};\frac{z^2}{2}\right)+\\Bz\, _1F_1\left(\frac{1-\nu}{2};\frac{3}{2};\frac{z^2}{2}\right)\Big\},
\end{multline}
where $z=\sqrt{2M_P\omega_P}(x-M/M_P^2)$ is the new dimensionless coordinate in the minisuperspace, $_1F_1(a;b;\zeta)$ is confluent hypergeometric
function, $\nu=\frac{1}{2}\left(\frac{M^2}{M_P^2} - 1 \right)$, $A$ and $B$ are two integration constants. 
Moreover,  from the asymptotic behavior of the confluent hypergeometric functions 
\begin{eqnarray}\label{PP2}
_1F_1(a;b;\zeta \rightarrow \infty)\simeq \Gamma(b)e^\zeta\zeta^{a-b}/\Gamma(a), \end{eqnarray}
we find that the  bracket $ \{ ... \}$ in Eq. (\ref{2-5}) brings the 
factor $e^{z^2/2}$,  which would dominate over the Gaussian exponential factor
explicit in (\ref{2-5}). In other words,
the wave function is square-integrable
  if the remaining factor 
provided from the bracket$ \{ ... \}$ in Eq. (\ref{2-5}),  vanishes.  This fixes the relative values of  $A$ and $B$, leading to
\begin{multline}\label{2-6}
\psi(z)=N 2^{\frac{\nu}{2}}e^{-\frac{z^2}{4}}\Big\{ \frac{_1F_1\left(-\frac{\nu}{2};\frac{1}{2};\frac{z^2}{2}\right)}{\Gamma(\frac{1-\nu}{2})}-\\z\frac{\sqrt{2}\,_1F_1\left(\frac{1-\nu}{2};\frac{3}{2};\frac{z^2}{2}\right)}{\Gamma(-\frac{\nu}{2})} \Big\},
\end{multline}
where $N$ is a normalization constant.
{Applying the boundary condition (\ref{2-4}) on the obtained wave function (\ref{2-6}) gives us
\begin{multline}\label{PP1}
    \frac{_1F_1\left(-\frac{\nu}{2};\frac{1}{2};(\frac{M}{M_P})^2\right)}{\Gamma(\frac{1-\nu}{2})}+\frac{2M\,_1F_1\left(\frac{1-\nu}{2};\frac{3}{2};(\frac{M}{M_P})^2\right)}{M_P\Gamma(-\frac{\nu}{2})}=0.
\end{multline}
The solution of the above equation for $M$, gives us the mass spectrum of the BH.

{We} can see that the spectrum of $M^2$ is not equally spaced for  small values of $\nu$.
On the other hand, for large values of  $\nu$ (which is defined by  $\nu=\frac{1}{2}\left(\frac{M^2}{M_P^2} - 1 \right)$) if we use the asymptotic relation (\ref{PP2}) in  (\ref{PP1})  we find
\begin{multline}
\frac{e^{2\nu+1}}{(2\nu+1)^{\frac{\nu+1}{2}}2^{\nu}\Gamma(-\nu)}=0\Rightarrow\Gamma(-\nu)=\pm\infty\Rightarrow \nu=n,
\end{multline}
}
which {provides us}
the mass spectrum 
\begin{eqnarray}\label{2-7}
M=M_P\sqrt{2n+1},
\end{eqnarray}
where $n$ is an integer and $n\gg1$. Note that the 
 above result assumes that $M \gg M_P$, i.e., $\nu \gg 1$ and is therefore 
essentially semiclassical.
To see this, let us apply the Bohr-Sommerfeld quantization rule to the Hamiltonian (\ref{1-9}). The classical turning points of (\ref{1-9}) are $x=0$ and $x=2M/M_P^2$. Hence, the conventional treatment of the Bohr-Sommer\-feld quantization rule  yields
\begin{multline}\label{NN1}
2\pi\left(n+\frac{1}{2}\right)=2\displaystyle\int_0^{\frac{2M}{M_P^2}}pdx\\
=2\displaystyle\int_0^{\frac{2M}{M_P^2}}\sqrt{M^2-M_P^4\left(x-\frac{M}{M_P^2}\right)^2}dx,
\end{multline}
which {allows to extract}
the mass spectrum (\ref{2-7}).
 Bekenstein \cite{Bekenstein}  firstly found a 
 similar mass spectrum. Generally, the proportionality constant for
 the square root of $n$ in (\ref{2-7}) is model dependent. Since then, several 
 authors \cite{Bekenstein,Kastrup} have used different arguments for a quantum BH spectrum of the type (\ref{2-7}). {Before proceeding, let us keep in mind equation (\ref{2-7}),  as this and others in this section will bear 
alterations that will be brought from  the intrinsic features of FQM, as
applied to the Schwarzshild BH 
(see next section)}.
 

{S. Hawking \cite{Hawking1,Hawking} showed in 1974 that due to quantum fluctuations, BHs emit black-body radiation, consistently with the corresponding entropy 
being }
one fourth of the event horizon area, namely $A=16\pi G^2M^2$.
Following  Refs. \cite{Mukhanov} and \cite{Xiang}, let us assume that {the} 
Hawking radiation of a massive BH
{where}
$M \gg M_P$ and  $n\gg 1$, is 
emitted when the BH system  spontaneously jumps from the state $n+1$ towards the closest lower state  level, i.e.,  $n$, as  described by (\ref{2-7}). 
{Let} us now denote the frequency of the emitted thermal radiation as $\omega_0$. Then
\begin{multline}\label{3-1}
  \omega_0=M(n+1)-M(n)\simeq\frac{M_P}{\sqrt{2n}}\\\simeq\frac{M_P^2}{M}\left[1+\frac{1}{2}\left(\frac{M_P}{M}\right)^2\right],  
\end{multline}
which agrees with the classical BH oscillation frequencies which {scales} as $1/M$.
We thus {find}
a BH to radiate with
a characteristic temperature $T\propto M_P^2/M$, matching the Hawking temperature.

The characteristic BH time   (the lifetime of the BH at the state $M(n+1)$ before decaying into the lower state $M(n)$) can be defined \cite{Xiang} as
\begin{eqnarray}\label{3-2}
\tau_n^{-1}=\frac{\dot M}{\omega_0}\simeq\frac{M\dot M}{M_P^2}\left[1-\frac{1}{2}\left(\frac{M_P}{M}\right)^2\right],
\end{eqnarray}
where $\dot M=dM/dt$ is the mass loss of the 
BH because of its evaporation;  in the second equality we used the definition of $\omega_0$ expressed in (\ref{3-1}).
As discussed in 
\cite{Mukhanov,Mukhanov2}, because of the interaction of the 
BH with the vacuum of the quantum fields, the width of  the
states, $W_n$, is not zero. The width of state $n$ can be estimated \cite{Mukhanov,Mukhanov2} 
as
\begin{eqnarray}\label{3-3}
W_n=\beta [M(n+1)-M(n)]=\beta\omega_0,
\end{eqnarray}
where $\beta\ll1$ is a numerical dimensionless factor. Then, by inserting (\ref{3-1}) into the uncertainty relation $W_n\tau_n\simeq 1$ and eliminating $\tau_n$ in resulting equation,  with
the assistance of (\ref{3-2}) we can find
\begin{eqnarray}\label{3-4}
\dot M=\frac{\beta M_P^4}{M^2}\left[1+\left(\frac{M_P}{M}\right)^2\right].
\end{eqnarray}

If we {further assume, on the one hand, that the origin of the}  Hawking radiation emerges from the highly blueshifted modes just outside the horizon, and, on the other hand,  {take} the BH as a black-body, then the radiated power is given by the Stefan–Boltzmann law \cite{Radiation,Radiation2}
\begin{eqnarray}\label{3-5}
\dot M=\sigma_SAT^4,
\end{eqnarray}
where $\sigma_S=\pi^2/60$ is the Stefan–Boltzmann constant and $A=16\pi M^2/M_P^4$ is the horizon area. Eliminating the mass loss of the BH, using Eqs.(\ref{3-4}) and (\ref{3-5}), gives us the effective temperature 
\begin{eqnarray}\label{3-6}
T=\left(\frac{\beta}{16\pi\sigma_S}\right)^\frac{1}{4}\frac{M_p^2}{M}\left[1+\frac{1}{4}\left(\frac{M_P}{M}\right)^2\right].
\end{eqnarray}
The BH entropy can then be expressed as
\begin{eqnarray}\label{3-7}
S=\int \frac{dM}{T}.
\end{eqnarray}
By choosing $\beta=1/15360\pi$ 
\cite{Correction}, 
Eqs. (\ref{3-6}) and (\ref{3-7}) {yields}
\begin{eqnarray}\label{3-8}
S=S_\text{B-H}-2\pi\ln{\left(S_\text{B-H}\right)}+const.,
\end{eqnarray}
where $S_\text{B-H}=4\pi GM^2$ is the  Bekenstein-{Hawking} entropy.
The logarithmic correction to the Bekenstein-Hawking entropy is obtained using other methods \cite{Correction} except that the overall factor $2\pi$ is model dependent.

\section{Fractional quantum mechanics and  Schwarzschild black hole thermodynamics}
\label{S4}

\indent

{In this section, 
we will establish and then investigate the alterations
brought from employing FQM features towards a 
Schwarzschild BH. This will provide broader 
expressions for some thermodynamic observables, specifically in terms explicitly dependent 
on the fractional parameter, $1<\alpha\leq 2$, since 
{the} limit $\alpha =2$ will conveys us to the standard case {(summarized}
in sections 2 and 3).}

Concretely, a 
fractional WDW equation for a  BH will be employed,  in a corresponding 
 minisuperspace. Our approach stems from 
Laskin's seminal papers \cite{Laskin1,Leskin1},
leading to an Hamiltonian, which includes a fractional kinetic term in terms of the quantum Riesz fractional operator; this  method has been extended to the WDW equation in Refs. \cite{FQ2,FQ3}.

\begin{figure}[ht]
\includegraphics[width=6cm]{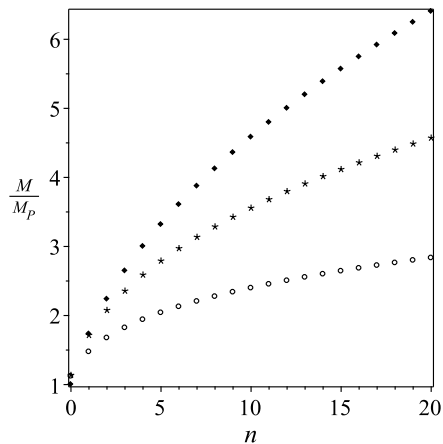}
\caption{Plot of the mass spectrum for a   Schwarzschild black hole as a function of quantum number $n$ for three values of $\alpha=1$ (circles), $\alpha=1.5$ (asterisks) and $\alpha=2$ (solid-diamonds). }\label{ppp1}
\centering
\end{figure}

The fractional extension of the WDW equation (\ref{2-2}) is given by 
\begin{multline}\label{4-1}
    \frac{1}{2}M_P^{1-\alpha}(-\Delta)^\frac{\alpha}{2}\psi(z)+\frac{1}{2}M_P^{\alpha-1}\omega_P^2z^\alpha\psi(z)=\\\frac{M^2}{2M_P}\psi(z),
\end{multline}
where $z=x-M/M_P^2$ is the new coordinate in the {$1$-dimensional} minisuperspace, $\Delta=d^2/dz^2$,  $(-\Delta)^\frac{\alpha}{2}$ is the Riesz fractional
derivative \cite{Rie,par1,par2,par3,Butzer} and $1<\alpha\leq2$. {Unfortunately, there is no known general solution,    explicitly bearing a dependence on $\alpha$  for the above fractional WDW equation. Therefore, we may resort to employ the Bohr-Sommerfeld quantization rule.}  By replacing $|p|=(-\nabla)^\frac{1}{2}$, the fractional WDW equation leads to the following relation
\begin{eqnarray}\label{w1}
\frac{M_P^{1-\alpha}}{2}|p|^\alpha+\frac{M_P^{\alpha-1}}{2}\omega_P^2z^\alpha=\frac{M^2}{2M_P},
\end{eqnarray}
which is, in fact, the fractional version of (\ref{1-9}). Hence, the classical turning points (where $|p|=0$) are $z=\pm(M^2/M_P^{2-\alpha})^\frac{1}{\alpha}$. 
Moreover, 
\begin{multline}
   2\pi\left(n+\frac{1}{2}\right)=\displaystyle\oint pdz=2\left(\frac{M}{M_P}\right)^\frac{4}{\alpha}\displaystyle\int_0^1\left(1-y^\alpha\right)^\frac{1}{\alpha}dy=\\\frac{2M^\frac{4}{\alpha}}{\alpha M_P^\frac{4}{\alpha}}B\left(\frac{1}{\alpha},1+\frac{1}{\alpha}\right), 
\end{multline}
where $y=(M_P^{\alpha+2}/M^2)^\frac{1}{\alpha}z$ and $B(a,b)$ is the beta function. Thus,
 the application of the 
standard Bohr-Sommerfeld quantization rule in this setting  gives
us the following  semi-classical mass {spectrum}
\begin{eqnarray}\label{4-2}
M=\left(\frac{\alpha\pi\Gamma(\frac{2}{\alpha})}{\Gamma(\frac{1}{\alpha})^2} \right)^\frac{\alpha}{4}M_P\left(n+\frac{1}{2}\right)^\frac{\alpha}{4}.
\end{eqnarray}
For $\alpha=2$ we recover  {spectrum (\ref{2-7}) as written in  the previous section}.
{Fig. (\ref{ppp1}) shows the BH mass spectrum for three  values of $\alpha$.} As we {show} 
in this figure, the mass of the BH {increases with $n$ but at a   faster rate}  for $\alpha=2$, {namely the standard case}.

\begin{figure}[ht]
\includegraphics[width=6cm]{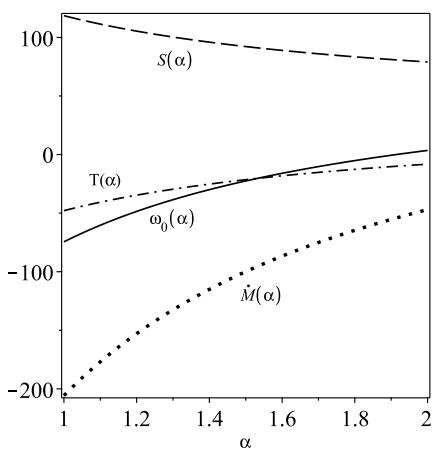}
\caption{{Logarithmic  plot of the fractional entropy (dash), $S(\alpha)$, the temperature (dash-dot), $T(\alpha)$,  the frequency of the emitted thermal radiation (line), $\omega_0(\alpha)$, and the mass loss (dot), $\dot M(\alpha)$ of the Schwarzschild black hole given by Eqs. (\ref{4-6}), (\ref{4-5}), (\ref{4-3}) and (\ref{4-4}) respectively,   with mass $M=10M_\odot$ as a function of $\alpha$.} }\label{ppp3}
\centering
\end{figure}

{Similarly to the previous section \ref{S3}, if we write the frequency of the  radiation emitted by the BH,  $\omega_0$ ,   we now obtain}
\begin{multline}\label{4-3}
    \omega_0(\alpha)=M(n+1,\alpha)-M(n,\alpha)\simeq\frac{\alpha BM_P}{4}n^{\frac{\alpha}{4}-1}\\
\simeq \frac{\alpha B^\frac{4}{\alpha}M_P^\frac{4}{\alpha}}{4M^{\frac{4}{\alpha}-1}}\left[1+\frac{1}{2}(1-\frac{\alpha}{4})\left(\frac{BM_P}{M}\right)^\frac{4}{\alpha}\right],
\end{multline}
where $B=\frac{\left(\alpha\pi\Gamma(\frac{2}{\alpha})\right)^\frac{\alpha}{4}}{{\Gamma(\frac{1}{\alpha})^\frac{\alpha}{2}}}$. Note that the energy of the emitted radiation is a function of the  fractional-order $\alpha$. We {then} find that for a massive BH,  where $M\gg M_P$, the energy of emitted radiation is {minimal} for $\alpha\rightarrow 1$, and it will { increases up to a maximum at  $\alpha=2$. Additional 
elements are conveyed within Fig. (\ref{ppp3}).}

{Interestingly,  the mass spectrum (\ref{4-2}) (which includes the standard case (\ref{2-7})) may induce observable signatures {that  gravitational waves \cite{GW} may inform about}.  
Eq. (\ref{3-1})  shows that the frequency of the emitted {Hawking} radiation  scales as $1/M$. Hence, if we put $M\simeq 10M_\odot-50 M_\odot$ as typical of the 
{BHs registered} by LIGO-Virgo, then we obtain from Eq. (\ref{3-1}) that  $f=\frac{\omega_0}{2\pi}=\mathcal O(10^2-10^4)$ Hz.  On the other hand, the fractional frequency of radiation $f(\alpha)=\frac{\omega_0(\alpha)}{2\pi}$, {whereas defined instead with} (\ref{4-3}), gives us a much  wider possibility for the frequencies. For example, with $M=10M_\odot$ the range $1<\alpha\leq 2$ yields $\mathcal O(10^{-74})\text{Hz}<f(\alpha)\leq \mathcal O(10^4)\text{Hz}$. Figure (\ref{ppp3}) shows how
$\omega_0(\alpha)$ depends on  $\alpha$.}

{To be more clear,
let us contemplate the {mass spectrum as in (\ref{4-2}) and further discuss it within a wider analysis as follows. Take the interaction} 
of a gravitational wave with a BH. Suppose the BH is initially at state $n_1$. Regarding the mass spectrum in (\ref{4-2}), we can also consider process whereby the  BH {could alternatively} absorb a gravitational wave whose frequency, $f_\text{GW}$ satisfies}
\begin{equation}\label{GW1a}
f_\text{GW}=\frac{\Delta M}{2\pi}=\frac{\alpha\pi\Gamma(\frac{2}{\alpha})\Delta n}{8\Gamma(\frac{1}{\alpha})^2M^{\frac{4}{\alpha}-1}},
\end{equation}
where $\Delta n=n_1-n_2$ and $n_2$ denotes the final state of the BH after absorption of the gravitational wave. 
As known in  classical GR, the effective potential that  describes the motion of a test particle has a maximum at the $3/2$ Schwarzschild radius, called photon sphere.
Essentially, this (potential barrier) screens the near-horizon region from external observers \cite{Potential}. 
The gravitational radiation proceeding towards 
 the horizon will be scattered at the horizon if its frequency does not match, e.g., (\ref{GW1a}). Otherwise,  radiation with the frequencies such as (\ref{GW1a})
in our example, would be absorbed. The previous assertions notwithstanding, if then we consider back-reaction quantum effects, BHs are not perfect absorbers of gravitational radiation and part of the radiation will be reflected from the horizon area. The reflected part of the radiation then interacts with the potential barrier at the photon sphere. Then, it will partially be transmitted, and the other portion will be reflected again back towards the horizon. This outcome generates a series of so-called gravitational-wave echoes \cite{Potential1}. This feature  could eventually be 
detected  either in the inspiral stage of BH binaries or during the last stages of BH relaxation following a merger. For {further} details {and detection methods}
please see \cite{GWnew}.  {Therefore, BHs may act as ``magnifying lenses'' in the sense that they could bring   
new features of the BH horizon-area within the realm of gravitational waves observations.}

{Furthermore}, 
using the formula of the characteristic time {of an evaporating  BH}, $\tau^{-1}=\dot M/\omega_0$, the uncertainty relation for width of states, $W_n\tau_n\simeq 1$ and the relation $W_n=\beta\omega_0$, we 
{compute}
\begin{eqnarray}\label{4-4}
\dot M=\frac{\alpha^2\beta B^\frac{8}{\alpha}M_P^\frac{8}{\alpha}}{16M^{\frac{8}{\alpha}-2}}\left[1+\left(1-\frac{\alpha}{4}\right)\left(\frac{BM_P}{M}\right)^\frac{4}{\alpha}\right],
\end{eqnarray}
which will increase as mass is lost. 
Inserting the above relation into the Stefan–Boltzmann law (\ref{3-5}), 
it gives the {corresponding} temperature of the BH\footnote{{{In our  paper we  assume that fractional calculus only induces alterations at quantum geometrodynamics level, and all the thermodynamic laws are essentially unchanged. We might extend the discussion to include fractional thermodynamics, though}. In this case, we should consider a fractional Stefan-Boltzmann law; see, e.g.,  \cite{Korichi}.}}
\begin{multline}\label{4-5}
    T=\\\left(\frac{\beta\alpha^2B^\frac{8}{\alpha}}{16^2\pi\sigma_S}\right)^\frac{1}{4}\frac{M_P^{\frac{2}{\alpha}+1}}{M^\frac{2}{\alpha}}\left[1+\frac{1}{4}\left(1-\frac{\alpha}{4}\right)\left(\frac{BM_P}{M}\right)^\frac{4}{\alpha}\right],
\end{multline}
\begin{eqnarray}
\beta=\frac{\Gamma(\frac{1}{\alpha})^4}{15(\alpha+2)^44^{\frac{4}{\alpha}-1}\Gamma(\frac{2}{\alpha})^2\pi^{\frac{4}{\alpha}+1}}.
\end{eqnarray}
Note that for $\alpha=2$ we will recover equations of the previous section.
{Moreover, {using (\ref{3-7}),} we find the fractional entropy of the BH to be}
\begin{multline}\label{4-6}
S=\\S_\text{B-H}^\frac{2+\alpha}{2\alpha}+\frac{\alpha(4-\alpha)(2+\alpha)4^\frac{\alpha+2}{2\alpha}\pi^\frac{3\alpha+2}{2\alpha}\Gamma(\frac{2}{\alpha})}{(2-\alpha)\Gamma(\frac{1}{\alpha})^2}\frac{1}{S_\text{B-H}^\frac{2-\alpha}{2\alpha}},\,\,\,\,\alpha\ne 2.
  \end{multline}

{To further contrast a BH with FQM features with
a standard Schwarzschild BH, let us examine Eqs.  (\ref{4-3}), (\ref{4-4}), (\ref{4-5}) and (\ref{4-6})   and   for a stellar BH with mass approximately
$10 M_\odot$. For the limiting $\alpha=2$ and $\alpha\rightarrow1$
values of the L\'evy index, the values of $S(\alpha)$, $T(\alpha)$, $\omega_0(\alpha)$ and $\dot M(\alpha)$ are given in Table (\ref{Table}).}
\begin{table}[h]
    \centering
    \begin{tabular}{|c|c|c|}
    \hline
         & $\alpha\simeq1$ & $\alpha=2$ \\
         \hline
         $S$ & $10^{118}$ & $10^{79}$\\
         \hline
        T& $10^{-48}$ K & $10^{-9}$ K \\
        \hline
$\dot M$ & $10^{-206}$ $kg/s$ & $10^{-47}$ $kg/s$ \\ 
\hline
$\omega_0$ & $10^{-74}$ Hz& $10^{4}$ Hz \\
\hline
    \end{tabular}
    \caption{{The fractional entropy, $S(\alpha)$, the temperature, $T(\alpha)$,   the mass loss, $\dot M(\alpha)$ and the frequency of the emitted thermal radiation, $\omega_0(\alpha)$ of the Schwarzschild black hole given by Eqs. (\ref{4-3}), (\ref{4-4}), (\ref{4-5})  and  (\ref{4-6}) respectively,   with mass $M=10M_\odot$. 
    }}
    \label{Table}
\end{table}
{Its first row, 
 together with Fig. (\ref{ppp3}), shows that the entropy of fractional BH can be much larger than in the standard Schwarzschild case.
The second row 
shows that a stellar BH temperature (with mass $10M_\odot$)
in fractional formalism is always less than 
$10^{-9} K$.
On the other hand, in the
extreme case  $\alpha\rightarrow1$, the temperature 
of a fractional BH is approximately zero: $T(\alpha\rightarrow 1) \simeq 10^{-48}$ K. 
Moreover, the 
manner the mass decay $\dot{M}(\alpha)$ proceeds is 
indicated by 
by the third row, 
in agreement with the temperature range: it  is {quite negligible}  for the extreme case, $\alpha\rightarrow 1$. Since the average temperature of the Universe at the present epoch is about 2.7 K, all stellar BHs with $M=10M_\odot$ and $1<\alpha\leq2$ are absorbing more matter and radiation than they emit {Hawking radiation} and will not begin to evaporate until the Universe has expanded and cooled below their corresponding temperature. Therefore, in our setting, the  BHs with fractional features could be almost eternal within $\alpha\rightarrow 1$}.

Let us now remark the following. 
Tsallis and Cirto \cite{Tsallis} investigated the entropy of a Schwarzschild BH,  using appropriate non-additive generalizations for
{$d-$di\-men\-sional} systems and suggested a generalized  entropy. In their study \cite{Tsallis}, a non-additive entropy is defined by
\begin{eqnarray}\label{4-7}
S_\delta=\sum_i^Np_i\left(\ln(\frac{1}{p_i})\right)^\delta,
\end{eqnarray}
 for a set of $N$ discrete states, where $\delta>0$ denotes the non-additivity
parameter and $p_i$ is a probability distribution \cite{Tsallis2}. For $\delta=1$ we recover the standard Boltzmann-Gibbs entropy. 
{Then,} as Tsallis and Cirto demonstrated \cite{Tsallis}, the generalized BH entropy can be written as
\begin{eqnarray}\label{4-8}
S_\text{Tsallis}\propto S_\text{B-H}^\delta.
\end{eqnarray}
If we identify the non-additivity parameter of {Tsallis and Cirto} as $\delta=\frac{2+\alpha}{2\alpha}$,    the entropy  (\ref{4-6}) 
retrieved from fractional quantum mechanical methods can be 
rewritten as
\begin{eqnarray}\label{4-9}
S=S_\text{B-H}^\delta+\frac{\kappa}{S_\text{B-H}^{\delta-1}},~~~~~~~1<\delta<\frac{3}{2},
\end{eqnarray}
where
\begin{eqnarray}
\kappa=\frac{\delta(2\delta-3)4^{\delta+1}\pi^{\delta+\frac{2}{2\delta-1}}\Gamma(2\delta-1)}{(\delta-1)(2\delta-1)^2\Gamma(\delta-\frac{1}{2})^2}.
\end{eqnarray}
This  shows that the leading term of the {FQM computed BH entropy} {can be expressed in terms of} 
the {Tsallis and Cirto} result for a BH. 
{Noteworthy,} 
for $1<\alpha\leq2$ the fractional entropy of a 
BH varies 
in the corresponding interval as
\begin{equation}\label{4-10}
S_\text{B-H}^\frac{3}{2}+\frac{36\pi^\frac{5}{2}}{S^\frac{1}{2}_\text{B-H}}<S\leq S_\text{B-H}-2\pi\ln{\left(S_\text{B-H}\right)}.
\end{equation}

{A natural question that 
arises is the meaning of the 
entropy relation, ranging from  Eqs. (\ref{4-6}) to (\ref{4-10})}. To 
assist in addressing this question, {let us first remind the fractal nature of the L\'evy path {\cite{Levy}} in FQM. As we know, in the Feynman path integral representation of quantum mechanics, the measure is generated by the process of the Brownian motion, and the fractal dimension of the Feynman's path is $d^{\text{(Feynman)}}_\text{fractal}=2$. {In addition,}, FQM is based on the fractional path integral \cite{Laskin1,Levy}. The L\'evy path integral corresponding to the SE (\ref{4-1}) is
\begin{eqnarray}
K_L(z_f,t_f|z_i,t_i)=\displaystyle\int_{z_i}^{z_f}\mathcal Dze^{-i\int_{z_i}^{z_f}V(z(t))dt},
\end{eqnarray}
where $V(z(t))$ is the potential as a functional of the L\'evy path and $z_i$ ($z_f$) denotes the initial (final) point. In this case, the fractional path integral measure is
\begin{multline}\label{NN11}
    \mathcal Dz(t)=\displaystyle\lim_{N\rightarrow\infty}dz_1...dz_{N-1}\times\\\displaystyle(iD_\alpha\varepsilon)^{-\frac{N}{\alpha}}\prod_{j=1}^{N}L_\alpha\left\{\left(\frac{1}{iD_\alpha\varepsilon}\right)^\frac{1}{\alpha}|z_j-z_{j-1}|\right\},
\end{multline}
where $D_\alpha=M_P^{1-\alpha}$, $\varepsilon=(t_f-t_i)/N$, $z_0=z_i$, $z_N=z_f$ and the L\'evy distribution function, $L_\alpha$, is indicated in terms of Fox’s $H$ function
\begin{multline}
    \left(D_\alpha t\right)^{-\frac{1}{\alpha}}L_\alpha\left\{\left(\frac{1}{D_\alpha t}\right)^\frac{1}{\alpha}|z| \right\}=\\\frac{1}{\alpha|z|}H_{2,2}^{1,1}\left[\left(\frac{1}{D_\alpha t}\right)^\frac{1}{\alpha}|z|\Big |^{(1,\frac{1}{\alpha}),(1,\frac{1}{2})}_{(1,1),(1,\frac{1}{2})}\right].
\end{multline}
The measure (\ref{NN11}) indicates that a length increment, $\Delta z=z_j-z_{j-1}$, and  a time increment, $\Delta t$, satisfy the fractional scaling relation
\begin{eqnarray}
\Delta z\propto D_\alpha^\frac{1}{\alpha}(\Delta t)^\frac{1}{\alpha}.
\end{eqnarray}
The above scaling relation implies that the fractal dimension of the L\'evy path 
is $d_\text{fractal}^{\text{(L\'evy)}}=\alpha$. 
{Thus,  fractional features within FQM 
could be interpreted as being  generated by a L\'evy 
stochastic process \cite{Levy} and suggesting a fractal 
structure}\footnote{According to Mandelbrot ``A fractal is by definition a set for which the Hausdorff-Besicovitch dimension strictly exceeds the topological dimension'' \cite{Benoit}. For example, the Hausdorff dimension of a regular Brownian surface is 2.79 and a triangular von Koch fractal Surface is $1+\log_2(3)=2.5849$
. }}. {Hence, it may be possible to theorize within the FQM framework herewith proposed, that the horizon of the BH may unveil a fractal 
structure whose relevance will be 
$\alpha$-dependent. 
In this context, 
let us rewrite the first term on the right-hand side of (\ref{4-6}) as follows
\begin{equation}
S_\text{B-H}^\frac{2+\alpha}{2\alpha}=\left(\frac{A}{4G}\right)^\frac{2+\alpha}{2\alpha}=\frac{A_\text{fractal}}{4G},
\end{equation}
by means of which we hence define  $A_\text{fractal}$.
Being more concrete, 
the entropy is 
proportional to a surface area,  but 
one explicitly given by }
\begin{equation}
    A_\text{fractal}=4L_p^2\left(\frac{A}{4G}\right)^\frac{2+\alpha}{2\alpha}.
\end{equation}
 Furthermore, the above expression {additionally} suggests a fractal dimension \cite{Benoit} of the BHs surface,  namely 
\begin{equation}
    D_\text{fractal}=\frac{2+\alpha}{\alpha},~~~~~2\leq D_\text{fractal}<3.
\end{equation}
{Moreover, 
equation (\ref{4-6}) will take the following form
\begin{multline}\label{frac1}
    S=\\\frac{A_\text{fractal}}{4G}+\kappa\left(\frac{A_\text{fractal}}{4G}\right)^{1-\frac{1}{2}D_\text{fractal}},~~~~2< D_\text{fractal}<3.
\end{multline}

Let us add 
further to the question set a few paragraphs above, namely on the meaning of entropy within Eqs. (\ref{4-6}) to (\ref{4-10}). 
By means of a very interesting paper, J. Barrow }recently conjectured that BHs might exist with such extremely wrinkled surfaces (referred to as a rough horizon), so that the event horizon would be a fractal surface
\cite{Barrow}. He proposed that the fractal area of the BH, $A_\text{fractal}$ is related to the ordinary area, $A=16\pi G^2M^2$, via
\begin{eqnarray}\label{NNN1}
A_\text{fractal}\propto A^{1+\frac{\triangle}{2}},~~~~0\leq\triangle\leq1,
\end{eqnarray}
where $\triangle=0$ is corresponding to the ordinary non-fractal horizon, and $\triangle=1$ to the most complex structure. Inserting the above area definition into entropy (\ref{4-6}) we find
\begin{eqnarray}
\label{NNN2}
S_\text{B-H}=\left(\frac{A}{4G}\right)^{1+\frac{\triangle}{2}}+\frac{\kappa}{\left(\frac{A}{4G}\right)^{\frac{\triangle}{2}}},~~~~0<\triangle<1,
\end{eqnarray}
where we defined $\triangle=2/\alpha-1$. Note that the extreme case of Barrow $\triangle=1$, ($\alpha\neq 1$), does not exist in our model. {Therefore, the above equation clarifies that the Barrow entropy can be obtained from within FQM.}

{ We can elaborate more }
on the  physical consequences of the above entropy (or equivalently from (\ref{4-6})) formula. If we assume that the number of degrees of freedom, $N$, in the horizon
is \cite{Komat}
\begin{eqnarray}\label{NN212}
N=4S,
\end{eqnarray}
then, combining {Eqs.  (\ref{4-5}),  (\ref{NNN2}) and (\ref{NN212}) plus some algebra}, we retrieve a modified equipartition theorem
\begin{eqnarray}
     M=\frac{1}{2}\left(1+\frac{\triangle}{2}\right)NT\Big\{1- \kappa_\triangle\left(\frac{T_P}{NT}\right)^{2(\triangle+1)}\Big\},
\end{eqnarray}
where
\begin{eqnarray}
\kappa_\triangle=\frac{\pi (\triangle+\frac{1}{2})\Gamma(\triangle+1)}{2^{2(\triangle+3)}\left( 1+\frac{\triangle}{2}\right)^{2(\triangle+1)}\Gamma(\frac{\triangle+1}{2})^2},
\end{eqnarray}
and $T_P$ is the Planck temperature.
The heat capacity, $C$
of a fractional 
BH can be computed from the expression
\begin{eqnarray}
C
=-\frac{(S')^2}{S''},
\end{eqnarray}
where a prime means a derivative relative to the BH mass, $M$.  Substituting (\ref{NNN2}) into the above definition gives
\begin{multline}
    C
    =\\-\frac{\triangle+2}{\triangle+1}\left(\frac{M}{M_P}\right)^{\triangle+2}\left[1-\frac{3\kappa_\triangle}{\triangle+2}\left(\frac{M_P}{M}\right)^{2\triangle+2} \right],
\end{multline}
which shows that for any value of $0<\triangle<1$, the heat capacity in our semiclassical model, $M\gg M_P$,  is negative and consequently the BH is unstable.

Thus, from the geometrical BH surface, we may  admit  that  there is  more to unveil. 
In particular, we may speculate that 
a behavior, suggesting  a 
broader dimensional dynamics
at the horizon, can be explored in the limit 
when L\'evy's parameter approaches $\alpha\rightarrow1$, supporting the conjecture proposed in \cite{Barrow}.
We  may, therefore, be allowed to speculate about the possibility 
that at some 
scale, a broader surface area of a BH can be considered, encompassing the standard value $16\pi G^2M^2$, because of the fractal structure of the horizon. In this context, let us recall Eq. (\ref{frac1}) 
regarding how the entropy varies and how the Bekenstein-Hawking expressions fits in. It brings an interesting perspective such that the  entropy of a BH (according to  FQM) could be 
 proportional to a power of its fractal surface,
depending on  the {choice} of $\alpha$.


\section{
Discussion and Outlook}

\indent

{The purpose in this paper was  to employ FQM 
to discuss specific  BH features.  It is now well demonstrated
that BHs can form \cite{Nobel} and  a set of  properties has been   widely described. Moreover,  BHs encompass scenarios 
 where strong gravity and quantum mechanics can  meet. New  insights need therefore to be proposed. Thus,   this paper was set up  to  explore 
 {whether} thermodynamic observables 
 might be altered if an additional ingredient is present (say, imported from a quantum mechanical framework)  and {if} it could  reveal itself in special BH  situations.} 
 
{Hence, we started by  briefly summarizing
 how concrete 
 thermodynamic properties can be retrieved from 
 a simple quantum mechanical BH (Schwarschild). This was done in sections 2 and 3. We then proceeded to an analysis employing a space-fractional derivative in a corresponding WDW equation. A broader 
 perspective about BH thermodynamics  within FQM  was {therefore presented} in section 4,  specifically in terms explicitly dependent 
on the fractional parameter, $1<\alpha\leq 2$. 
We retrieved altered expressions for physical quantities that were then 
 contrasted with those of the standard case ($\alpha=2$) {and summarized} in sections 2 and 3.}

{{Then in section 4 we started 
by elaborating on 
the 
 mass spectrum  of a BH  and how it varies according to $\alpha$.} 
Whereas the frequency of gravitational waves generated by a {standard BH 
is proportional to the inverse 
mass, in a fractional scenario the frequency would instead allow a dependence as }
$1/M^{\frac{4}{\alpha}-1}$. As pointed out in \cite{GWnew}, the horizon behaves as a filter of gravitational waves with a set of absorption lines. {Admitting the possibility of} the fractal structure of the horizon, the horizon radius is different from the Schwarzschild radius \cite{Barrow}, and the effective radius is {wider}, $r=2MG(1+\epsilon)$,  in which $\epsilon\ll 1$ depends on how 
the surface is { correspondingly altered. Any subsequent  gravitational waves may bear
a delay due to the time it takes the signal to transit from the  horizon (with such possible fractal structure) to the photon sphere. This delay time scales logarithmically with $\epsilon$ \cite{GWnew} or, as we suggested herein, the  fractality of the horizon. }}

{In addition, {we have shown how} the entropy, the temperature, and the mass loss of the BH are highly sensitive to the value of $\alpha$.
The Bekenstein-Hawking entropy of supermassive black holes \cite{Egan} is of the order of $10^{104}$ and  the corresponding fractional entropy of the supermassive BHs with {$\alpha \sim 1$ }
will be $10^{156}$. If we extend a FQM perspective 
to the cosmic event horizon (CEH), then we can compute that 
$S_\text{CEH}\propto (GH^2)^\frac{-(\alpha+1)}{2\alpha}\simeq 10^{183}$. This assumption 
has been interestingly employed to discuss the late-time acceleration of the Universe \cite{Barrow2}.
Furthermore, we retrieved a  formula for the Schwarzschild BH in which the entropy 
is a polynomial function of its area.
{We then {explained} that the surface area of 
{such}  BH can 
exceed the standard value $16\pi G^2M^2$ of a BH,  as suggested by a possible  fractal structure. 
In this context, Tsallis and Cirto's formalism \cite{Tsallis} must be mentioned, constituting  an extension of  Boltzmann--Gibbs statistical theory, representing a non-extensive, i.e., a non-additive entropy\footnote{{This definition of entropy encompasses the conventional properties of positivity, equiprobability, concavity, and irreversibility.  In addition, it has been favorably used in many different physical systems. For example, we can mention the L\'evy-type anomalous diffusion \cite{TT1}, turbulence in a pure-electron plasma \cite{TT2}, and gravitational systems \cite{TT3}.}} {We have indicated herewith  that from a FQM perspective we can alternatively,  obtain}  Tsallis and Cirto's  results  without committing to the extended entropy features. {In addition, we {proposed} 
that the inherent fractal nature of L\'evy paths leads 
to the extension of Barrow entropy formula \cite{Barrow}.  Barrow's modified entropy formally coincides with Tsallis and Cirto’s definition; notwithstanding, the involved foundations and interpretation of physical results are different. This is not a surprising result because, over the last several years, it has been shown that a fractal structure in properties of physical systems could be a possible origin for non-additive statistics \cite{Fractaltsallis}.  }}}

{Having summarized and discussed our results, it is also pertinent 
to suggest  an outlook of subsequent directions to prospect. A particular line in our sight is 
to explore BH entropy  and information loss  
within FQM. We plan to extend the context of paper \cite{Marto} 
with the assistance of space-fractional derivative operators in the corresponding equations ther\-ein. The results in \cite{Marto} 
indicated that  for the  final stage of a BH evaporation,   which included back reaction from Hawking radiation, we get a strong entangled state (between a mimicked BH and the radiation). So, given how the entropy may vary with $\alpha$, what would FQM add within a suitable set up about information and BHs?}

{
It would also be of interest to investigate BHs within FQM and in a Brans-Dicke or even a scalar-tensor theory, where more parameters are allowed to juggle with. Or employ other dynamical configurations instead, beyond the  Schwarzschild case. It is possible using other fractional derivatives as well, besides space-fra\-ctional. Most likely, the use of numerical methods would be mandatory.}

\section*{Acknowledgments}
This research work was supported by Grant No. UID / MAT / 00212 / 2020  
and COST Action CA18108 (Quantum gravity phenomenology in the multi-me\-ssen\-ger approach). {The authors are grateful to M. Ortigueira for his kind feedback  and sharing reading suggestions on the scope of fractional calculus}.

\end{document}